\documentclass{article}
\usepackage{romp}

\usepackage{graphicx}
\usepackage{amsfonts}
\usepackage{amssymb,amsmath}

\begin{document}

\title{Trace distance and linear entropy of qubit states: Role of initial qubit-environment correlations
}
 \author{Jerzy Dajka \\ Institute of Physics, University of Silesia, 40-007 Katowice, Poland  \\ e-mail: jerzy.dajka@us.edu.pl \\[2ex]
         Jerzy {\L}uczka \thanks{The work  supported by   the  NCN Grants N202 052940 (J. {\L}.),  UMO-2011/01/B/ST6/07197 and 2011/01/N/ST3/02473 (J. D.)}
                      \\ Institute of Physics, University of Silesia, 40-007 Katowice, Poland  \\ e-mail: jerzy.luczka@us.edu.pl}

\maketitle

\begin{abstract} 
The  role of initial qubit--environment correlations on trace 
distance between two qubit states is studied in the framework  of non--Markovian pure dephasing.  
 The growth of  mixedness of reduced state quantified by  linear entropy  is shown to be related to the degree of initial qubit--environment correlations.     
\end{abstract}



%
\section{ Introduction}

In quantum mechanics, a state of a system is defined by a density matrix $\rho$. Nowadays, the state   can be completely determined by the procedure which is called quantum tomography. 
Due to the fundamental limitations
related to the Heisenberg uncertainty principle and the no-cloning
theorem \cite{zurek},  one cannot perform an arbitrary sequence of
measurements on a single system without inducing on it a back-action of some
sort. On the other hand, the no-cloning theorem forbids to create a perfect copy
of the system without already knowing its state in advance. Thus, there is no
way out, not even in principle, to infer the quantum state of a single system
without having some prior knowledge on it \cite{ariano}.
However, it  is possible to reconstruct the unknown quantum state of a system when an ensemble of 
identical copies are available in the same state, so that a series of ideal measurements 
can be performed on each copy, for more information see e.g. \cite{tomo,kwiat}. 

Manipulation  on quantum states in atomic, molecular and optical systems is an important problem  in contemporary physics research,
with many types of experiments aimed at applications ranging from metrology
 to quantum computation,  quantum cryptography and 
quantum-state engineering. 
 It naturally arises the question on the relation between states before and after the manipulation and the extend to which states are similar or different. For example,  are the output states  more  distinguishable or less distinguishable than the input states; do they behave in a similar way or not.    The answer in not simple because it depends on a distinguishability measure.  In literature one can find many  examples of distinguishability measures which a part are based on the notion of a distance $D[\rho_1, \rho_2]$ between two states 
$\rho_1$ and $\rho_2$. The distance can be quantified by a metric which has the following properties:  
\\
(i) non-negativity $D[\rho_1, \rho_1] \ge 0$, \\
(ii) identity of indiscernibles $D[\rho_1, \rho_2]=0$ 
if and only if $\rho_1=  \rho_2$, \\
(iii) symmetry 
$D[\rho_1, \rho_2] = D[\rho_2, \rho_1]$,  \\
(iv) the triangle inequality 
$D[\rho_1, \rho_2] \le D[\rho_1, \rho_3] + D[\rho_3, \rho_2]$. 

 In fact, the first condition is implied by the others. 
Because there are infinitely many metrics, the problem arises which metric is proper \cite{gil}.
The most popular distance measures include the  trace distance, Hilbert-Schmidt distance,  Bures distance, Hellinger distance  and Jensen-Shannon divergence  \cite{gil,luo,majtey,dodonov,nielsen,karol}. 

Majority of description of a general state-transformation  is given in terms of the so-called
 quantum operation (also named as quantum dynamical map, quantum process  or quantum channel) ${\cal E}$, i.e. by 
a linear, trace-preserving (more generally: trace non-increasing)  completely positive  map. 
Let us recall that the quantum operation $\cal E$ defined on the whole space of operators $\rho$ on the Hilbert space   is contractive with respect to a given distance $D[\rho_1, \rho_2]$  if
\begin{eqnarray}\label{contr}
D[{\cal E}(\rho_1),  {\cal E}(\rho_2) ] \le D[\rho_1, \rho_2].
\end{eqnarray}
It implies that two quantum states do not become more distinguishable under the action of
quantum operation. 

Real quantum systems are typically open, i.e. they interact with environment. Dynamics of such systems is not unitary and in general it is even not described by quantum operation.   To understand it, let us assume that the total system $S$ + environment $E$ is closed and its unitary dynamics is determined by the  
total Hamiltonian   
\begin{eqnarray}
H=H_S\otimes {\mathbb{I}}_E+H_I+{\mathbb{I}}_S \otimes H_E, 
\end{eqnarray}
where $H_S$ is the Hamiltonian of the system $S$, $H_E$ is the Hamiltonian of the environment $E$ and 
$H_I$ describes  the system-environment interaction. The operators 
${\mathbb{I}}_S$   and ${ \mathbb{I}}_E$  are   identity operators (matrices) in corresponding Hilbert spaces of the system $S$ and the environment $E$, respectively.
Let the initial state $\varrho$  of total system $S +E$ is determined by the density operator   
\begin{eqnarray}\label{init1}
\varrho =  \varrho^{SE}(0). 
\end{eqnarray} 
Then the state of the system  $S$ at time $t > 0$ is determined by the reduced dynamics, 
\begin{eqnarray}\label{red}
\rho(t) = \mbox{Tr}_E \{\varrho^{SE}(t)\}= \mbox{Tr}_E \{ U(t) \  \varrho^{SE}(0) \  U^\dagger(t)\}, \quad U(t)=\exp[-iHt/\hbar].  
\end{eqnarray}
Because the partial trace $\mbox{Tr}_E$ is a quantum operation, it defines the  trace preserving positive map $S+E \to S$: 
\begin{eqnarray}\label{pres}
\varrho^{SE}(0) \to  \Lambda_t \varrho^{SE}(0)= \mbox{Tr}_E \{ U(t) \  \varrho^{SE}(0) \  U^\dagger(t)\}. 
\end{eqnarray}
%
In the case when the initial state is  a non-correlated state, i.e. when  it is a product state 
\begin{eqnarray}\label{product} 
\varrho^{SE}(0) =\rho_S \otimes\omega^E,  
\end{eqnarray} 
where $\rho_S$ is an arbitrary  initial state of the system $S$ and $\omega^E$ is a {\it fixed}   initial state of the 
environment $E$, the relation 
\begin{eqnarray}\label{noncor}
\rho_S \to  \Phi_t \rho_S = \mbox{Tr}_E \{U(t) \ \rho_S\otimes\omega^E \  U^\dagger(t)
\}
\end{eqnarray}
defines a trace preserving  quantum operation  $S \to S$ (into itself)  which is also called a completely positive quantum dynamical map.   Let  ${\cal E} = {\Phi_t}$ is  such that $\rho^{\Phi}(t) = {\Phi_t} \rho(0)$, where  $\rho(0)$ is an initial state of the system $S$. Then contractivity (\ref{contr})  means that
\begin{eqnarray}\label{contr2}
D[\rho^{\Phi}_1(t) , \rho^{\Phi}_2(t)  ] \le D[\rho_1(0), \rho_2(0)] 
\end{eqnarray}
for any two initial states $\rho_1(0)$ and $\rho_2(0)$ of the system $S$. 
As a consequence, the distance between two system states cannot increase in time and the  distinguishability of any states  can not increase above its initial value.  

Now, let the initial state $\varrho^{SE}(0)$ is not a product state. It means that the system $S$ is initially correlated with its environment $E$.  
Let  ${\cal E} = {\Lambda_t}$ is  such that $\rho^{\Lambda}(t) = {\Lambda_t} \varrho^{SE}(0)$, where  $\varrho^{SE}(0)$ is an initial state of the total system $S$.  Then contractivity (\ref{contr})  means that
\begin{eqnarray}\label{contr3}
D[\rho^{\Lambda}_1(t) , \rho^{\Lambda}_2(t)  ] \le D[\rho_1^{SE}(0), \rho_2^{SE}(0)]. 
\end{eqnarray}
Note that in the left hand side of this relation there are two states  $\rho^{\Lambda}_1(t)$ and $\rho^{\Lambda}_2(t)$ the system $S$ while in the right hand side  
there are two states $\rho_1^{SE}(0)$ and $\rho_2^{SE}(0)$ of the total system $S+E$. 
In this  case,  one cannot say whether the distance between two states of the system $S$ 
decreases or not because the relation (\ref{contr3}) does not imply the inequality 
\begin{eqnarray}\label{contr3b}
D[\rho^{\Lambda}_1(t) , \rho^{\Lambda}_2(t)  ] \le D[\rho_1(0), \rho_2(0)],   
\end{eqnarray}
where
\begin{eqnarray}\label{correl}
\rho_i(0) = \mbox{Tr}_E \{\varrho^{SE}_i(0)\} \quad i=1, 2
\end{eqnarray}
is the initial reduced state of the system $S$.

The formal relation 
\begin{eqnarray}\label{red}
\rho_S=\mbox{Tr}_B \{\varrho^{SE}(0)\} \to W_t \rho_S = \mbox{Tr}_E \{ U(t) \  \varrho^{SE}(0) \  U^\dagger(t)\}
\end{eqnarray}
is not generally a quantum operation. Moreover, $W_t$  is not even a map because many different  $\varrho^{SE}(0)$  reduce to the same $\rho_S$ and for the same $\rho_S$ one can obtain several  different $W_t \rho_S$. Therefore in general  the inequality (\ref{contr3b}) need not be fulfilled in the case of initially correlated state of $S+E$.

As an example of the distance measure let us recall the {\it trace distance} defined by the relation
\begin{eqnarray}\label{trace}
D_T[\rho_1, \rho_2]=\frac{1}{2}\mbox{Tr} \sqrt{(\rho_1-\rho_2)^2} 
\end{eqnarray}
which is limited to the unit interval,
\begin{eqnarray}
0\le D_T[\rho_1, \rho_2] \le 1.  \nonumber
\end{eqnarray}
From the Ruskai theorem \cite{ruskai} follows that the quantum operation $\cal E$ is a 
contraction with respect to the trace distance. In such a case,  the relations (\ref{contr2}) and  
 (\ref{contr3}) are satisfied when  distance $D=D_T$ is the  trace distance. 
Therefore the trace distance between two states of the system $S$ cannot increase in time when the system $S$ is initially non-correlated with its environment  
 and the  distinguishability of any system states  can not increase above an initial value. 
In the case when the system $S$ is correlated with  the environment $E$, 
 one cannot say whether the distance between two states of the system $S$ 
decreases or not because the relation (\ref{contr3b}) does not hold in a general case. 
 
  From the above it follows that  the trace distance $D_T[\rho_1(t), \rho_2(t)]$ 
  between two states $\rho_1(t)$ and $\rho_2(t)$ 
  of the system $S$ can grow above its initial value $D_T[\rho_1(0), \rho_2(0)]$  only 
  in two cases: \\
  (A) when  the system $S$ is initially correlated with  the environment $E$   or \\
  (B) when the system $S$ is initially  non-correlated with  the environment $E$ and  in the relation  (\ref{noncor}) there are two different initial states $\omega^E_1$ and $\omega^E_2$ of the environment $E$, i.e. when  
\begin{eqnarray}\label{different}
\rho_1(t) = \mbox{Tr}_B \{U(t) \ \rho\otimes\omega^E_1 \  U^\dagger(t) \}, \nonumber\\
\rho_2(t) = \mbox{Tr}_B \{U(t) \ \rho\otimes\omega^E_2 \  U^\dagger(t) \}.
\end{eqnarray}

We should remember that contractivity of reduced dynamics is not a universal feature but depends on chosen  metric and therefore  decrease or increase of distances between two states can depend on the metric \cite{daj1}.   Contractivity of quantum evolution can break down when the system $S$  is initially correlated with its environment \cite{pech, romero2004} and implications of 
 such correlations have been  studied in various context \cite{romero2004, correla}. 
Examples of an exact reduced dynamics which fail contractivity with respect to the {\it trace distance} are studied in Refs. \cite{daj,breu3,daj1}. Two experiments on initial system-environment correlations have recently been conducted in optical systems \cite{exper}.

In the paper, we study the role of initial system-environment states on the trace distance of states and linear entropy for the reduced dynamics. 
In Sec. 2,   we define a quantum open system. It is a two-level system (qubit) interacting with an  infinite bosonic environment. We consider a pure dephasing interaction between the qubit and the
environment \cite{defaz} and  we  ignore the energy decay of the qubit. This
assumption is reasonable since in some cases the phase coherence decays much
faster than the energy.  We derive the exact  reduced dynamics of the qubit  for a particular initial correlated qubit-environment state. Properties of time evolution of the distance 
between  two states of the qubit are demonstrated in Sec. 3. In the same section we analyze the  mixedness of reduced state quantified by entropic measure.  Finally,  Sec. 4  provides summary and some conclusions.

 \section{Qubit dephasingly coupled to infinite bosonic environment}

In this section, we consider the same model as in our previous papers \cite{daj,daj1}. For the readers convenience and to keep the paper self-contained, we repeat all necessary definitions and introduce  notations. The model consists of  a qubit $Q$ (two-level system)   coupled to its  environment $E$ and we limit our considerations to the case when the process of energy dissipation  is negligible and only pure dephasing 
 is acting as the mechanism responsible for decoherence of the qubit dynamics \cite{defaz}.
Such a decoherence mechanism can  be described  by the total Hamiltonian  (with $\hbar=1$)
\begin{eqnarray}\label{ham}
H= H_Q \otimes  {\mathbb{I}}_E +{\mathbb{I}}_Q  \otimes H_E + S^z \otimes H_I, \\
H_Q = \varepsilon S^z, \quad
H_E=\int_0^\infty d\omega \, h( \omega) a^\dagger(\omega)a(\omega), \\
H_I=\int_0^\infty d\omega  \left[ g^*(\omega)a(\omega) +g(\omega)a^\dagger(\omega)\right],
\end{eqnarray}
where  $S^z$  is the z-component of the spin operator and is represented by the diagonal matrix $S^z =diag[1,-1]$ of elements $1$ and $-1$. The parameter  $ \varepsilon$ is the  qubit energy splitting,   ${\mathbb{I}}_Q$   and ${ \mathbb{I}}_E$  are   identity operators (matrices) in corresponding Hilbert spaces of the  qubit $Q$ and the environment $E$, respectively.
The operators $ a^\dagger(\omega)$ and $a(\omega)$ are the bosonic creation and annihilation operators, respectively.  The real-valued spectrum function $h(\omega)$  characterizes the environment. The coupling is described  by the function $g(\omega)$ and  the function
 $ g^*(\omega)$ is   the complex conjugate to   $g(\omega)$.
The Hamiltonian (\ref{ham}) can be rewritten in the  block--diagonal structure \cite{fidel},
\begin{eqnarray} \label{H1}
H=diag[H_{+}, H_{-}],  \quad
H_\pm =  H_B  \pm   H_I  \pm  \varepsilon { \mathbb{I}}_B.
\end{eqnarray}
As an example, we assume that at the initial time $t=0$ the composite
wave function $|\Psi(0)\rangle$  of $S+E$  is given  by the expression 
\begin{eqnarray}\label{ini}
|\Psi(0)\rangle =b_+|1\rangle \otimes |\Omega_0\rangle +b_-|-1\rangle \otimes
|\Omega_{\lambda}\rangle.
\end{eqnarray}
The states $|1\rangle$ and $|-1\rangle$ denote the excited and
ground state of the qubit, respectively. The non-zero complex numbers $b_+$ and $b_-$ are chosen  in such  a way that the condition 
$|b_+|^2 + |b_-|^2 =1$ is satisfied.  The state
 $|\Omega_0\rangle$   is  the  ground  (vacuum) state of the environment and
\begin{eqnarray}\label{omega1}
|\Omega_{\lambda} \rangle = C_{\lambda}^{-1} \, \left[(1-\lambda)|\Omega_0\rangle +\lambda |\Omega_f\rangle  \right],
\end{eqnarray}
where  $|\Omega_f\rangle =D(f) |\Omega_0\rangle$
is  the coherent state. The  displacement (Weyl) operator $D(f) $  reads \cite{brat}
\begin{eqnarray}\label{displacement}
D(f)=\exp\left\{\int_0^\infty d\omega \left[ f(\omega)a^{\dagger}(\omega) -  f^*(\omega)a(\omega)\right]\right\}
\end{eqnarray}
 for an arbitrary square--integrable function $f$.
The constant  $C_{\lambda}$ normalizes the state (\ref{ini}) and is given by the expression
\begin{eqnarray}\label{C}
C_{\lambda}^2=(1-\lambda)^2+\lambda^2+2\lambda(1-\lambda)
Re \langle \Omega_0|\Omega_f\rangle,
\end{eqnarray}
where ${Re}$ is a real part of the scalar product $\langle \Omega_0|\Omega_f\rangle $ of two states in the environment Hilbert space.  
The  correlated  initial state (\ref{ini}) is in the form  similar to that in  Ref.  \cite{daj}. 
The  parameter $\lambda\in[0,1]$  controls  the   strength of initial correlations  of the qubit with  environment.  For $\lambda=0$ the qubit  and the environment are initially uncorrelated while for  $\lambda=1$ the correlation is most prominent   for the assumed class of initial states.

The initial wave function (\ref{ini})  of the isolated system  $S+E$ evolves unitarily according to the 
Hamiltonian (\ref{ham}) and reads 
\begin{eqnarray}  \label{ewol}
|\Psi(t)\rangle =b_+|1\rangle \otimes |\psi_+(t)\rangle +b_-|-1\rangle \otimes
|\psi_-(t)\rangle,
\end{eqnarray}
where
\begin{eqnarray}  \label{ewolB}
|\psi_+(t)\rangle &=& \exp(-i H_+ t)   |\Omega_0\rangle, \nonumber\\
|\psi_-(t)\rangle &=& \exp(-i H_- t)   |\Omega_{\lambda}\rangle.
\end{eqnarray}
 The state of the total system is a pure state and the corresponding density  matrix 
 $\varrho(t) = |\Psi(t) \rangle \langle\Psi(t)|$. In turn, the partial trace  $\mbox{Tr}_E$ over the environment degrees of freedom yields
 the density  matrix  $\rho_{\lambda}(t) =  \mbox{Tr}_E \varrho(t)$ of the qubit.
In the base $\{|1\rangle, |-1\rangle\}$, it takes the  matrix form 
\begin{eqnarray}\label{ro}
\rho_{\lambda}(t)=\left(\begin{array}{cc} |b_+|^2 & b_+b_-^* \, A_{\lambda}(t) \\
b_+^*b_- \,   A_{\lambda}^*(t) & |b_-|^2 \end{array}  \right),
\end{eqnarray}
where the decoherence factor $A_{\lambda}(t)$ is given by  
\begin{eqnarray}\label{A}
 A_{\lambda}(t)= C_{\lambda}^{-1} \,e^{-2i\varepsilon t - r(t)}  \left[1-\lambda+\lambda
e^{-2i\Phi(t) +  s(t)}  \right],
\end{eqnarray}
and  \cite{fidel}
\begin{eqnarray}\label{r}
r(t) &=& 4\int_0^\infty d\omega g_h^2(\omega) \left[1-\cos(\omega t) \right], \quad \quad\quad \quad\quad \\
\label{s}
s(t) &=& 2\int_0^\infty d\omega g_h(\omega)f(\omega)\left[1-\cos(\omega t)\right]   \quad \nonumber\\
&-&  \frac{1}{2} \int_0^\infty d\omega f^2(\omega), 
\end{eqnarray}
%
where $g_h(\omega) = g(\omega)/\omega$ and
\begin{eqnarray}\label{Fi}
\Phi(t) =  \int_0^\infty d\omega g_h(\omega)f(\omega) \sin(\omega t).
\end{eqnarray}
Without loss of  generality we have assumed in Eqs. (\ref{r})-(\ref{Fi})  that the  functions $g(\omega)$ and $f(\omega)$ are real valued and the energy spectrum function $h(\omega) = \omega$.

Now, let us consider the second class of initial states of the total system. 
We assume that the system-environment initial  state is mixed and given by  the product state:
\begin{eqnarray} \label{prod1}
\varrho_p(0)=\rho_\lambda(0)\otimes |\Omega_0\rangle\langle\Omega_0|, 
\end{eqnarray}
where 
\begin{eqnarray} \label{product}
\rho_{\lambda}(0) =  \mbox{Tr}_E  |\Psi(0) \rangle \langle\Psi(0)|
\end{eqnarray}
is the marginal qubit state and  $ |\Psi(0) \rangle$ is defined by Eq. (\ref{ini}). 
From the relation 
\begin{eqnarray}\label{proev}
\rho_p(t) =     \mbox{Tr}_E \{  \mbox{e}^{-iHt} \ \rho_\lambda(0)\otimes |\Omega_0\rangle\langle\Omega_0|  \   \mbox{e}^{-iHt} \}
\end{eqnarray}
we obtain the reduced dynamics of the qubit  in  the case of the initial uncorrelated qubit-environment state.  It can be expressed in the matrix form as:
\begin{eqnarray}\label{ro0}
\rho_{p}(t)=\left(\begin{array}{cc} |b_+|^2 & b_+b_-^* \, A_{\lambda}(0)A_{0}(t) \\
b_+^*b_- \,   A^*_{\lambda}(0)A^*_{0}(t) & |b_-|^2 \end{array}  \right).
\end{eqnarray}
Properties of the trace distance between qubit states subjected to reduced dynamics (\ref{ro}) and (\ref{ro0}) are presented in  the next section. 
We stress that that  initial states (\ref{ini}) and (\ref{prod1}) of the total systems are  different because the initial environmental  states are different. However,  the reduced {\it initial states of the qubit are the same} in both cases.


\section{Properties of trace distance and linear entropy}

Our model is still incomplete. We have to consider some models for 
 the spectral density $g_h^2(\omega)$ of the environment, see Eqs.(\ref{r})-(\ref{Fi}). 
 We assume that for low  frequencies it exhibits power--like  frequency dependence and the frequency scale characterizes the cut-off frequency. 
 An example of such a function is taken in the form \cite{leg}
\begin{eqnarray}\label{J}
g_h^2(\omega)=\alpha  \, \omega^{\mu-1}\exp(-\omega/\omega_c),
\end{eqnarray}
where $\alpha >0$ is the qubit-environment coupling constant, $\omega_c$ is a cut-off  frequency and $\mu>-1$ is the "ohmicity" parameter:
the case $-1<\mu <0$  corresponds to the sub--ohmic, $\mu =0$ to the ohmic and $\mu >0$  to super--ohmic environments, respectively.  As it follows from our previous study \cite{daj}, only in the case of super--ohmic environment, the trace distance can increase. Therefore below we analyze only this regime.

The second function we have to specify is 
 the function $f(\omega)$  in Eq. (\ref{displacement}) which  determines the coherent state  of the boson environment. 
We can choose any integrable function but for convenience  we take the function 
\begin{eqnarray}\label{f}
f^2(\omega)&=&\gamma \, \omega^{\nu-1}\exp(-\omega/\omega_c).
\end{eqnarray}
There is no deeper physical justification for it and the only reason for our choice is possibility to calculate explicit formulas for  the functions in Eqs. (\ref{r})-(\ref{Fi}). As a result one gets
\begin{eqnarray}\label{LL}
r(t)=4\mathcal{L}(\alpha,\mu,t),   \quad \quad  \quad     \\
s(t)= 2\mathcal{L}(\sqrt{\alpha \gamma},(\mu+\nu)/2,t)-\frac{1}{2} \gamma\Gamma(\nu)\omega_c^\nu,
 \end{eqnarray}
\begin{eqnarray}\label{el}
\mathcal{L}(\alpha,\mu,t) =\alpha\Gamma(\mu) \omega_c^\mu\left\{1-\frac{\cos\left[\mu\arctan(\omega_c t) \right] }{(1+\omega_c^2t^2)^{\mu/2}}\right\}, 
\end{eqnarray}
\begin{eqnarray}\label{phi}
\ \ \ \nonumber \\
\Phi(t) = \sqrt{\alpha \gamma}\; \Gamma\left(\kappa\right)
 \omega_c^{\kappa}
\;  \frac{\sin\left[\frac{\mu+\nu}{2} \arctan(\omega_c t) \right] }{(1+\omega_c^2t^2)^{\kappa/2}},  
\end{eqnarray} 
where $\kappa =(\mu+\nu)/2$
and $\Gamma(z)$ is the Euler gamma function.

\begin{figure}[htpb]
\includegraphics[width=0.5\linewidth, angle=270]{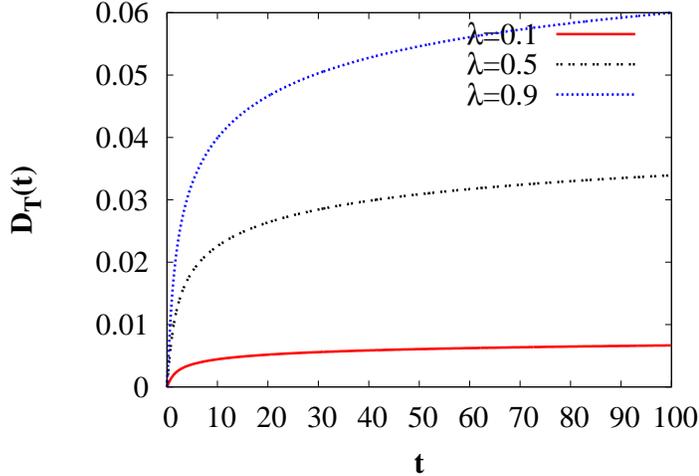}
\caption{(Color online) Time evolution of the  trace distance 
$D_T(t)\equiv D_T[ \rho_{\lambda}(t), \rho_p(t)]$ between two qubit states  (\ref{ro}) and (\ref{ro0}) for selected values of the correlation parameter $\lambda$.  Time is in unit of $\omega_c$, the dimensionless coupling $\alpha \omega_c^{\mu} = 0.01$ and   $\gamma \omega_c^{\nu}=0.05$.    The remaining parameters are: $\varepsilon =1,   \mu=0.01$, $\nu=0.2$
and $|b_+^{(1)}|^2 = |b_+^{(2)}|^2 =1/2$.
}
\label{fig1}
\end{figure}

\begin{figure}[htpb]
\includegraphics[width=0.5\linewidth, angle=270]{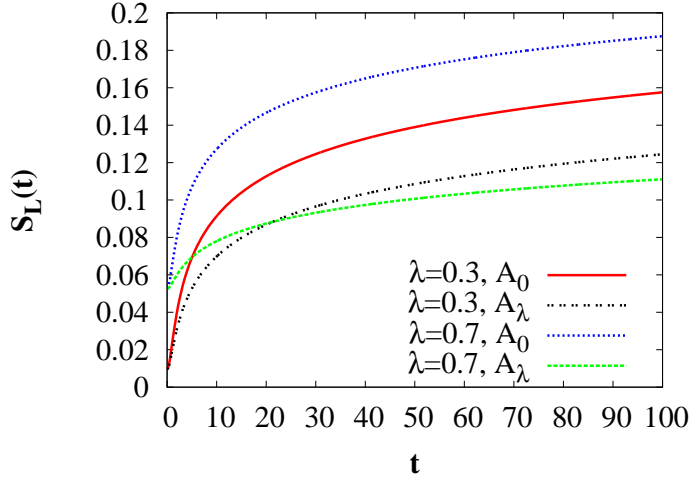}
\caption{(Color online) Time evolution of  the linear entropy of  two qubit states (\ref{ro}) indicated  by $A_\lambda$ and (\ref{ro0}) 
indicated  by $A_0$  for selected values of the correlation parameter $\lambda$.  Time is in unit of $\omega_c$, the dimensionless coupling $\alpha \omega_c^{\mu} = 0.01$ and   $\gamma \omega_c^{\nu}=0.05$.    The remaining parameters are: $\varepsilon =1,   \mu=0.01$, $\nu=0.2$
and $|b_+^{(1)}|^2 = |b_+^{(2)}|^2 =1/2$.
}
\label{fig2}
\end{figure}
We analyze  the distance between two qubit states for  the case when two corresponding initial states 
of the total system  are given by Eqs. (\ref{ini}) and (\ref{prod1}), respectively. It is the case when at initial time $t=0$ the first state  of the qubit is correlated with its environment and the second  state is non-correlated. 
  In Fig. \ref{fig1} we present time evolution of the  trace distance 
  $D_T(t)= D[ \rho_{\lambda}(t), \rho_p(t)]$ between two qubit states  (\ref{ro}) and (\ref{ro0}) for selected values of the correlation parameter $\lambda$. The first observation is the monotonic increase of the distance as time grows and the distance saturates in the long-time limit.   The second observation is that increase of  the correlation parameter $\lambda$  enhances distinguishability of two states and the distance between two  qubit states  grows.

 Quantum dynamics of an open system results in growing (or at least non--lowering) 'mixedness' of the reduced state. Equivalently, one may discuss the problem of the entropy production resulting from the dephasing process.  Here, as a simplest measure of the information loss due to system-environment  interaction we adapt the  linear entropy
\begin{eqnarray}
\label{sl}
S_L[\rho]=1-\mbox{Tr}(\rho^2)
\end{eqnarray}
of any reduced states $\rho$. It takes the form 
\begin{eqnarray}\label{S1}
S_L[\rho_\lambda] = 1- \left[|b_+|^4+|b_-|^4+2|b_+|^2|b_-|^2|A_\lambda(t)|^2\right] 
\end{eqnarray}
for the state (\ref{ro}) and 
\begin{eqnarray}\label{S2}
S_L[\rho_p] = 1- \left[|b_+|^4+|b_-|^4+2|b_+|^2|b_-|^2|A_\lambda(0) 
A_0(t)|^2\right] 
\end{eqnarray}
for the state (\ref{ro0}). 

The linear entropy  can range between zero, corresponding to a completely pure state, and $1/2$  corresponding to a completely mixed state. 
In Fig. {\ref{fig2},  we compare time evolution of the linear entropy  for both types of models Eqs. (\ref{ro},\ref{ro0}). It is seen that the difference 
$S_L[\rho_p] - S_L[\rho_\lambda]$ is greater when $\lambda$ is greater. It means that for any time $t>0$  the mixedness of the reduced state is smaller when the qubit is initially correlated with its environment in comparison to the case when the initial   state 
is uncorrelated. Moreover,  if the degree of correlation grows the mixedness decreases. 
The results exemplified in Fig. {\ref{fig2} allow to conclude that 
\begin{eqnarray} \label{St}
S_L[\rho_p]\ge S_L[\rho_\lambda]. 
\end{eqnarray}
As the  expression for both $S_L[\rho_\lambda]$ and $S_L[\rho_p]$ are known,  it  can be proven using straightforward  analytic methods. 
Indeed, from (\ref{S1}) and (\ref{S2}), it follows that (\ref{St}) holds true if $|A_\lambda(t)|^2 \ge |A_\lambda(0) A_0(t)|^2$. 
In turn, it is equivalent to the requirement $s(t) \ge s(0)$ which is true as follows from Eq. (\ref{s}).  

%
\begin{figure}[htpb]
\includegraphics[width=0.5\linewidth, angle=270]{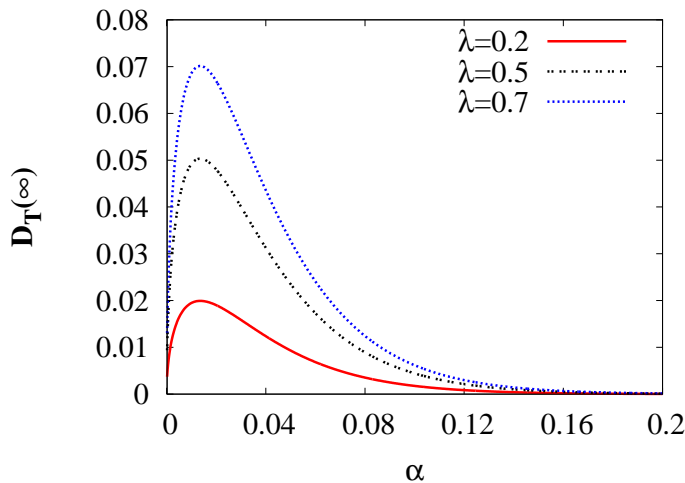}
\caption{(Color online) Long--time characteristics of the  trace distance between two qubit states  (\ref{ro}) and (\ref{ro0}) as a function of the qubit-environment coupling constant $\alpha$ in  (\ref{J}) for selected values of the correlation parameter $\lambda$.  Time is in unit of $\omega_c$,   $\gamma \omega_c^{\nu}=0.05$.    The remaining parameters are: $\varepsilon =1,   \mu=0.01$, $\nu=0.2$
and $|b_+^{(1)}|^2 = |b_+^{(2)}|^2 =1/2$.
}
\label{fig3}
\end{figure}

\begin{figure}[htpb]
\includegraphics[width=0.5\linewidth, angle=270]{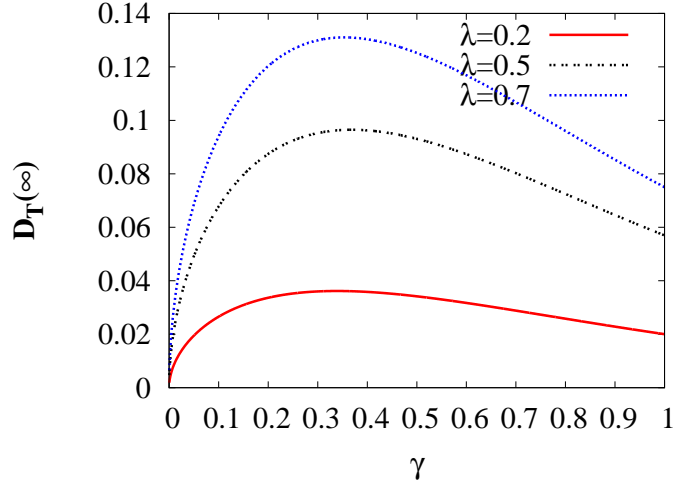}
\caption{(Color online) Long--time characteristics of the  trace distance between two qubit states Eqs. (\ref{ro}) and (\ref{ro0}) as a function of the amplitude $\gamma$ Eq,(\ref{f}) for selected values of the correlation parameter $\lambda$.  Time is in unit of $\omega_c$,   the dimensionless coupling $\alpha \omega_c^{\mu} = 0.01$.    The remaining parameters are: $\varepsilon =1,   \mu=0.01$, $\nu=0.2$
and $|b_+^{(1)}|^2 = |b_+^{(2)}|^2 =1/2$.
}
\label{fig4}
\end{figure}

\begin{figure}[htpb]
\includegraphics[width=0.5\linewidth, angle=270]{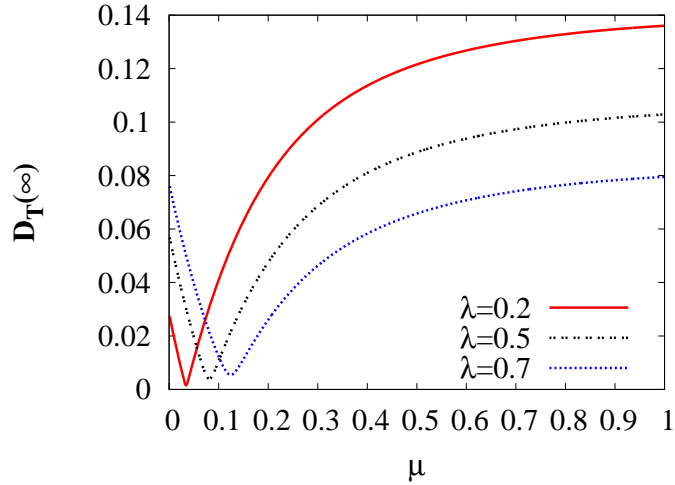}
\caption{(Color online) Long--time characteristics of the  trace distance between two qubit states  (\ref{ro}) and (\ref{ro0}) as a function of the 'ohmicity parameter' $\mu$ in Eq.(\ref{J}) for selected values of $\lambda$.  Time is in unit of $\omega_c$, the dimensionless coupling is kept fixed: $\alpha \omega_c^{\mu} = 0.01$ and   $\gamma \omega_c^{\nu}=0.05$.    The remaining parameters are: $\varepsilon =1$,  $\nu=0.2$ and $|b_+^{(1)}|^2 = |b_+^{(2)}|^2 =1/2$.}
\label{fig5}
\end{figure}
Now, we analyze the long time limit of the  trace distance between two qubit states,  
$D_T(\infty) = \lim_{t\to \infty} D_T[\rho_{\lambda}(t), \rho_p(t)]$. The dependence of $D_T$ on the 
 qubit-environment coupling constant $\alpha$, which occurs in the spectral density  (\ref{J}), 
 exhibits a bell-shaped extremum  and then the optimal coupling  exists which maximizes the trace distance, see Fig. \ref{fig3}.  The maximum is the most pronounced for a strong initial correlations. For weak and strong coupling, the distance tends to zero.  
In Fig. \ref{fig4}, we depict the influence of the parameter $\gamma$ which characterizes the coherent state of environment. One can also observe a  maximum of the distance. Finally, in Fig. \ref{fig5}, 
we investigate  how 
the spectral properties of the environment, encoded in the 'ohmicity' parameter $\mu$, influences on the distance. The most important observation is the occurrence of minimum in  distance for a particular value of $\mu$. The strong superomic environment is more desired for  distingushability of qubit states. 

\section{Summary}

 According to Ref. \cite{breu3}, an increase of the distance can be interpreted
in terms of the exchange of information between the system and its environment. If the distance
increases over its initial value, information which is locally inaccessible at the initial time is transferred to the open system. This transfer of information enlarges the distinguishability of the open-system states which suggests various ways for the  experimental detection of initial correlations. The trace metric is one of the most important measures of a distance between states in quantum information processing. Moreover, it has a physical interpretation as a measure of state distinguishability. 

With this work, we  have studied the role of initial qubit-environment correlations and analyzed two characteristics: trace distance between two qubit states and the linear entropy.  We have demonstrated  that the trace distance exhibits a rich diversity of  characteristics and is sensitive to selected parameters like coupling strength of the qubit and environment.  In particular,
depending on the chosen parameter regime, we can identify optimal regimes where the distance is locally maximal and distinguishability of the qubit states can be  maximal.  Moreover, the trace distance increases  as time grows and  saturates in the long-time limit.   The increase of  the correlation parameter $\lambda$  enhances distinguishability of two states and the distance between two  qubit states  grows. The results on the trace distance presented here extend findings of Refs. \cite{daj} and \cite{daj1},  where the  initial state of the composite system was  always  a pure state and the initial state of the environment was a mixed state.  Here we include the case of a mixed state of the composite system and a pure environment state. Here and there, the initial state of the qubit is mixed. 

We  have also considered a linear entropy of the reduced state which  monotonically grows as time grows (the reduced state is more and more mixed). 
The impact of initial correlations on the linear entropy is also crucial:  For any time $t>0$  the mixedness of the reduced state  
(i.e. how much the initial state is far from being pure)
is smaller when the qubit is initially correlated with its environment and if the degree of correlation grows the mixedness decreases. 
This work could be continued to extend future  theoretical studies. In
particular, it would be interesting to study other classes of  initial correlations and their impact on 
distance properties of qubit states and other characteristics of the quantum open systems.

\end{document}